\documentclass[conference]{IEEEtran}
\usepackage{graphicx}
\usepackage{amsmath}
\usepackage{lipsum}
\usepackage{amsthm}
\usepackage[strings]{underscore}
\usepackage{amsfonts}
\usepackage{caption}
\usepackage{siunitx}
\usepackage{amssymb}
\usepackage{subcaption}
\usepackage{placeins}
\usepackage{lettrine}
\usepackage{multicol, blindtext}
\usepackage{bbm}
\usepackage{cite}
\usepackage{color}
\usepackage{algorithm}
\usepackage{algpseudocode}
\usepackage{authblk}

\makeatletter

\newcommand{\Rmnum}[1]{\expandafter\@slowromancap\romannumeral #1@}
\newenvironment{psmallmatrix}
  {\left[\begin{smallmatrix}}
  {\end{smallmatrix}\right]}
\newtheorem{theorem}{Theorem}

\newtheorem{proposition}{Proposition}

\makeatletter
\def\BState{\State\hskip-\ALG@thistlm}
\makeatother
\ifCLASSINFOpdf
\else
\fi

\begin{document}
%
\title{Age of Information With Prioritized Streams: When to Buffer Preempted Packets?}
\author[*]{Ali Maatouk}
\author[*]{Mohamad Assaad}
\author[$\dagger$]{Anthony Ephremides}
\affil[*]{TCL Chair on 5G, Laboratoire des Signaux et Syst\`emes, CentraleSup\'elec, Gif-sur-Yvette, France }
\affil[$\dagger$]{ECE Dept., University of Maryland, College Park, MD 20742}
\maketitle

\begin{abstract}
In this paper, we consider $N$ information streams sharing a common service facility. The streams are supposed to have different priorities based on their sensitivity. A higher priority stream will always preempt the service of a lower priority packet. By leveraging the notion of Stochastic Hybrid Systems (SHS), we investigate the Age of Information (AoI) in the case where each stream has its own waiting room; when preempted by a higher priority stream, the packet is stored in the waiting room for future resume. Interestingly, it will be shown that a ``no waiting room" scenario, and consequently discarding preempted packets, is better in terms of average AoI in some cases. The exact cases where this happen are discussed and numerical results that corroborate the theoretical findings and highlight this trade-off are provided.

\end{abstract}


%
\IEEEpeerreviewmaketitle

\section{Introduction}
\lettrine{T}{he} Age of Information is a new metric that has been introduced in \cite{6195689} and quantifies the notion of data freshness at a monitor about a physical process of interest. This notion is considered of broad interest in numerous applications where information sources generate time-stamped status updates that are sent through the network towards their intended receiver. Some non-exclusive examples of such applications include environmental monitoring and vehicular networks \cite{5597912}\cite{5307471}. As the main goal in these applications is to maximize data freshness at the monitor's side, a surge in papers investigating the minimization of the age of information in a variety of environments can be witnessed. 

In \cite{6195689}, the authors studied the average AoI in the standard First-Come-First-Served (\textbf{FCFS}) queuing models: M/M/1, M/D/1 and D/M/1. Subsequent to the work in \cite{6195689}, the authors in \cite{6875100} showed that the management of packets can further minimize the AoI. The aforementioned reference also introduced the new notion of average \emph{peak} AoI. Investigation of the AoI in sampling problems where sources generate packets at will was done in \cite{8000687}. Interestingly, it was shown that, in general, a zero wait sampling policy is far from being optimal in minimizing the average AoI. The AoI has also attracted attention in energy harvesting environments (e.g. \cite{2018arXiv180202129A}). With the majority of the early work on the AoI revolving around single-hop networks, multi-hop scenarios have recently attracted research attention 
\cite{8262777,8445981,8006593} where for example, in the latter, it was proven that the Last-Come-First-Served (\textbf{LCFS}) discipline at relaying nodes minimizes the average AoI. Scheduling with the goal of minimizing the average AoI of a broadcast network (e.g. cellular network) has also been examined considerably in the litearture (e.g. \cite{8006590}). As the majority of the AoI scheduling literature focused on Orthogonal Multiple Access (\textbf{OMA}) schemes where only a single user access the channel, the authors in \cite{2019arXiv190103020M} investigated the potentials of Non-Orthogonal Multiple Access (\textbf{NOMA}) in minimizing the average AoI. Knowing that the AoI is mainly of interest in machine type communications where distributed scheduling schemes are required, the authors in \cite{2019arXiv190100481M} provided an optimal back-off scheme to minimize the average AoI in a Carrier Sense Multiple Access (\textbf{CSMA}) environment.

Priority based queuing has been extensively studied in the queuing theory literature. This is a natural consequence of the vast amount of real-life scenarios where information streams are assigned different priorities based on their sensitivity. A simple example of such scenarios is when critical safety data is being sent along with non-safety related data. With the AoI being a relatively new metric, the literature investigating this metric in priority based queuing scenarios is limited \cite{2018arXiv180805738Z,2018arXiv180104067N,2018arXiv180511720M,8437591}. The AoI in a multicast scenario where two different priority groups exist was studied in \cite{2018arXiv180805738Z}. The average AoI of two different priority streams, each having its own queuing discipline, has been investigated in \cite{2018arXiv180511720M}. Recently, the authors in \cite{8437591} investigated a scenario where $N$ streams, each with a different assigned priority, share the same service facility. The aforementioned reference studied the case where a higher priority stream preempts the service of a lower priority packet and consequently discards it. Our paper seeks to answer the following natural question: instead of discarding the preempted packet, should it be kept in its own waiting room for resume after the high priority packet finishes being served? At a first glance, the answer appears to be a trivial ``Yes" as discarding a packet due to an arrival of a higher priority packet will badly influence the age of the lower priority stream and would lead to the server being utilized less. However, it will be shown in the sequel that this is not always the case and there are scenarios where a ``no waiting room" setting yields a lower AoI. It is worth noting that, from a technical point of view, introducing the individual waiting room for each stream hugely complicates the analysis as the age of the buffered packets after preemption have to be constantly tracked unlike the case where they are simply discarded upon arrival of a higher priority stream as in \cite{8437591}. To proceed with our analysis, we first leverage the notion of Stochastic Hybrid Systems (\textbf{SHS}) to find a closed form of the average AoI of each stream. As our scenario is general for any number of streams $N$ and as it includes individual waiting room for each stream, the SHS analysis is involved and the closed forms of the average age of each stream are obtained after elaborate steps. Armed with these closed forms, we investigate in which cases it is better to opt for a ``no waiting room" setting rather than the latter. Numerical results that corroborate the theoretical findings and highlight this trade-off are provided.
 
The paper is organized as follows: Section \Rmnum{2} describes the system model in question. Section \Rmnum{3} presents the theoretical results on the average AoI of each stream. Section \Rmnum{4} provides the numerical results that utilize the theoretical findings while Section \Rmnum{5} concludes the paper.

\section{System Model}
We consider in our paper $N$ information streams sharing a common service facility. The streams are considered of different priority, i.e., a higher priority stream will always preempt the service of a lower priority stream. This scenario is of broad practical interest: a simple example can be a sensor gathering different information such as temperature, humidity and accelerometer data where the temperature data is considered to be the most crucial and therefore have a priority over the other streams. Other examples include a vehicular network where crucial safety data are given priority over non-safety related data that aim to improve the traveling experience. In the sequel, we consider that the transmission time of each of these packets to be exponentially distributed with an average rate of $\mu$. We also assume that the packets arrival of the streams to be exponentially distributed with an average rate of $\lambda$. 
In our paper, we consider the case where the preempted packet of a lower priority stream is stored in its own single buffer space. Any new arrival of this low priority stream will replace the old packet in the buffer space. Once the higher priority stream's packet finishes being served, the service of the low priority stream is resumed\footnote{It is worth mentioning that a new arrival belonging to a particular stream, when the server is already serving a packet of this particular stream, will preempt its service and the new packet will take its place. This setting is motivated by the fact that a preemptive M/M/1/1 scenario was shown to minimize the average age in the case of exponential transmission time \cite{8006593}.}. 

We first start by introducing the notion of average AoI. The instantaneous age of information at the receiver (monitor) of stream $k$ at time instant $t$ is defined as:
\begin{equation}
\Delta_k(t)=t-U_k(t)
\end{equation}
where $U_k(t)$ is the time-stamp of the last successfully received packet by the receiver side of stream $k$. Clearly the evolution of the age will depend on the arrival process of each stream, along with the transmission time and the interaction with the higher priority streams. Therefore, the ultimate goal consists of minimizing the total average age of the system that is defined as:
\begin{equation}
\overline{\Delta}^{WQ}=\sum_{k=1}^{N}\overline{\Delta}^{WQ}_k=\sum_{k=1}^{N}\lim_{\tau\to\infty}\frac{1}{\tau}\int_{0}^{\tau}\Delta^{WQ}_k(t)dt
\end{equation}
where $WQ$ refers to ``With Queues" to signal that there is an available buffer space (waiting room) for each of the individual streams.
\section{Theoretical Analysis}
\subsection{Preliminaries on SHS}
As the graphical approach to find the average AoI of the system (e.g. \cite{6195689}) can be challenging in lossy systems where packets might be preempted and discarded \cite{2018arXiv180307993Y}, we tackle this problem by leveraging the notion of SHS \cite{DBLP:journals/corr/YatesK16}. The SHS approach involves modeling the system 
through the states $(q(t),\boldsymbol{x}(t))$ where:
\begin{itemize}
\item $q(t)\in\mathbb{Q}$ is a process of discrete nature that aims to capture how the system evolves as events take place (e.g. a packet arrival, a packet finishing being served etc.)
\item $\boldsymbol{x}(t)\in\mathbb{R}^{2N}$ is a process of continuous nature that represents the evolution of the age of each stream $k$ at the monitor along with the packet in the system of stream $k$.
\end{itemize}
More specifically, we have:
\begin{equation}
\boldsymbol{x}(t)=[x_0(t),x_1(t),\ldots,x_{2N-2}(t),x_{2N-1}(t)]
\end{equation}
where:
\begin{equation*}
\begin{cases}
x_{2k}(t)& \text{Age of stream}\: $k+1$\: \text{at the monitor} \: 0\leq k\leq N-1\\
x_{2k+1}(t)& \text{Age of stream} \: $k+1$ \:  \text{system's packet} \: 0\leq k\leq N-1\\
\end{cases}
\label{Stationary}
\end{equation*}
In the remainder of this section, we \emph{briefly} present the main idea of SHS and we refer the readers to \cite{DBLP:journals/corr/YatesK16} for more details.

We start first by describing $q(t)$: this process is a Markovian one that we can fully characterize graphically using a Markov chain $(\mathbb{Q},\mathbb{L})$. In this chain, $\mathbb{Q}$ represents the set of vertices for which $q$ belong while $\mathbb{L}$ represents the set of transitions\footnote{A subtle difference between the aforementioned chain and a typical continuous-time Markov chain is the fact that, unlike the latter, the chain $(\mathbb{Q},\mathbb{L})$ may include self-transitions where $\boldsymbol{x}(t)$ experiences a drop but the discrete process $q(t)$ keeps the same value.} between those states. More specifically, each transition $l\in\mathbb{L}$ is a directed edge $(q_l,q'_{l})$ with a transition rate $\lambda^{(l)}\delta_{q_l,q(t)}$. The multiplication by the Kronecker delta function ensures that this transition may only happen when $q(t)=q_l$.
For each state $q$, we define the incoming and outgoing transitions sets respectively as:
\begin{equation}
\mathbb{L}'_q=\{l\in\mathbb{L}: q'_{l}=q\} \quad \mathbb{L}_q=\{l\in\mathbb{L}: q_{l}=q\}
\end{equation}
The interest in SHS originates from the fact that the discrete process transitions will result in a reset mapping in the continuous process. In other words, as a transition $l$ happens, the discrete process changes to $q'_{l}$ and a reset in the continuous process $\boldsymbol{x'}=\boldsymbol{x}\boldsymbol{A_l}$ is witnessed. The matrix $\boldsymbol{A_l}\in\mathbb{R}^{2N}\times\mathbb{R}^{2N}$ is called the transition \emph{reset maps}. Moreover, to fully capture the evolution of $\boldsymbol{x}(t)$, we point out that in each state $q\in\mathbb{Q}$, the continuous process evolves through the following differential equation $\dot{\boldsymbol{x}}=\boldsymbol{b}_q$ where $b^k_q$ is a binary element that is equal to $1$ if the age process $x_k$ increases at a unit rate when the system is in state $q$ and is equal to $0$ if it keeps the same value.

In order to use SHS to calculate the average age of the system, we define the following quantities for all states $q\in\mathbb{Q}$:
\begin{equation}
\pi_{q}(t)=\mathbb{E}[\delta_{q,q(t)}]=P(q(t)=q)
\end{equation}
\begin{equation}
\boldsymbol{v}_{q}(t)=[v_{q0}(t),\ldots,v_{q2N}(t)]=\mathbb{E}[\boldsymbol{x}(t)\delta_{q,q(t)}]
\end{equation} 
where $\pi_{q}(t)$ is the Markov chain's state probabilities and $\boldsymbol{v}_{q}(t)$ denotes the correlation between the age process and the discrete state of the system $q$. The Markov chain $q(t)$ is supposed to be ergodic and, consequently, we define the steady state probability vector $\overline{\boldsymbol{\pi}}$ as the solution to the following set of equations:
\begin{equation}
\overline{\pi}_{q}(\sum_{l\in\mathbb{L}_q}\lambda^{(l)})=\sum_{l\in\mathbb{L}'_q}\lambda^{(l)}\overline{\pi}_{q_l} \quad q\in\mathbb{Q}
\end{equation}
\begin{equation}
\sum_{q\in\mathbb{Q}}\overline{\pi}_{q}=1
\end{equation}
As it was proven in \cite{DBLP:journals/corr/YatesK16}, in this case,  the correlation vector $\boldsymbol{v}_{q}(t)$ converges to $\overline{\boldsymbol{v}}_{q}$ such that:
\begin{equation}
\overline{\boldsymbol{v}}_{q}(\sum_{l\in\mathbb{L}_q}\lambda^{(l)})=\boldsymbol{b}_q\overline{\pi}_{q}+\sum_{l\in\mathbb{L}'_q}\lambda^{(l)}\overline{\boldsymbol{v}}_{q_l}\boldsymbol{A_l} \quad q\in\mathbb{Q}
\label{solutionv}
\end{equation}
Taking that into account, we can come to a conclusion that $\mathbb{E}[x_{2k}]=\lim\limits_{t \to +\infty} \mathbb{E}[x_{2k}(t)]=\lim\limits_{t \to +\infty}\sum\limits_{q\in\mathbb{Q}}\mathbb{E}[x_{2k}(t)\delta_{q,q(t)}]=\sum\limits_{q\in\mathbb{Q}}\overline{v}_{q2k}\:\:\forall k\in\{0,\ldots,N-1\}$.\\
Based on the presented results from \cite{DBLP:journals/corr/YatesK16} and knowing that our aim is to calculate the average age at the
monitor of each stream $k$, we provide the following theorem
that summarizes all what have been previously detailed.
\begin{theorem}
When the Markov chain $q(t)$ is ergodic and admits $\boldsymbol{\overline{\pi}}$ as stationary distribution, if we can find a solution for eq. (\ref{solutionv}), then the average age at the monitor of stream $k \:\:\forall k\in\{1,\ldots,N\}$ is:
\begin{equation}
\overline{\Delta}_k=\sum_{q\in\mathbb{Q}}\overline{v}_{q2(k-1)}
\end{equation}
\label{theomhem}
\end{theorem}
\subsection{Average age calculation}
In order to simplify the average age calculations, we forgo studying the age process of all streams simultaneously. Instead, we examine the perspective of a stream of interest $k$ that can be preempted by $0\leq i\leq N-1$ higher priority streams. Based on this, we can define the discrete states $\mathbb{Q}=\{0,1,2,\ldots,i\}$ where $q(t)=0$ if the server is serving the stream of interest and $q(t)=j$, with $1\leq j\leq i$, if there are $j$ packets of higher priority streams that have to be served before the server is able to work on the stream of interest. The continuous-time process is defined as $\boldsymbol{x}(t)=[x_{0}(t),x_{1}(t)]$ where $x_{0}(t)$ is the age of the stream of interest $k$ at the monitor at time $t$ and $x_{1}(t)$ is the age of the packet that is being served (or awaiting in the buffer to be served upon service completion of higher priority streams) of stream $k$ at time $t$. To further simplify the average age calculation of the stream of interest, we suppose that if there are no packets of stream $k$ in the system, a ``fake" update packet is considered to be available. A fake update packet is defined as a packet that has the same time-stamp as the previously received packet by the monitor. Introducing the fake update will not change the average age calculation but will provide mathematical benefits; it allows the reduction of the state space of $\mathbb{Q}$ since the availability of a packet of stream $k$ in the system does not need to be monitored. Our goal is to apply Theorem \ref{theomhem} to find the vectors $\overline{\boldsymbol{v}}_{q}=[\overline{v}_{q0},\overline{v}_{q1}]\:\: \forall q\in\mathbb{Q}$ that will enable us to find the average age of the stream $k$. In order to do this, we present in the following table the transitions between the discrete states and the reset maps they induce on the age process $\boldsymbol{x}(t)$:
\begin{center}
\begin{tabular}{cccccc}
$l$ & $q_l\rightarrow q'_{l}$ & $\lambda^{(l)}$ & $\boldsymbol{xA_l}$ & $\boldsymbol{A_l}$  &  $\boldsymbol{v}_{q_l}\boldsymbol{A_l}$ \\
\hline
$1$ &  $0\rightarrow0$  & $\lambda$  & $[x_0,0]$ & $\begin{psmallmatrix}
    1 & 0  \\
    0& 0
\end{psmallmatrix}$   &  $[v_{00},0]$ \\
      $2$  &  $1\rightarrow1$  & $\lambda$ & $[x_0,0]$ & $\begin{psmallmatrix}
    1 & 0  \\
    0& 0
\end{psmallmatrix}$   & $[v_{10},0]$ \\
            & $\vdots$ & $\vdots$ & $\vdots$ & $\vdots$  &  $\vdots$\\
     $i+1$ & $i\rightarrow i$ & $\lambda$ & $[x_0,0]$ & $\begin{psmallmatrix}
    1 & 0  \\
    0& 0
\end{psmallmatrix}$   &  $[v_{i0},0]$ \\
 $i+2$ & $0\rightarrow1$  & $i\lambda$ & $[x_0,x_1]$ & $\begin{psmallmatrix}
    1 & 0  \\
    0& 1
\end{psmallmatrix}$  &  $[v_{00},v_{01}]$\\
    & $\vdots$ & $\vdots$ & $\vdots$ & $\vdots$  &  $\vdots$\\
  $2i+1$ &  $i-1\rightarrow i$  & $\lambda$  & $[x_0,x_1]$ & $\begin{psmallmatrix}
    1 & 0  \\
    0& 1
\end{psmallmatrix}$   &  $[v_{(i-1)0},v_{(i-1)1}]$ \\
  $2i+2$ &  $i\rightarrow i-1$  & $\mu$  & $[x_0,x_1]$ & $\begin{psmallmatrix}
    1 & 0  \\
    0& 1
\end{psmallmatrix}$   &  $[v_{i0},v_{i1}]$ \\
& $\vdots$ & $\vdots$ & $\vdots$ & $\vdots$  &  $\vdots$\\
$3i+1$ &  $1\rightarrow 0$  & $\mu$  & $[x_0,x_1]$ & $\begin{psmallmatrix}
    1 & 0  \\
    0& 1
\end{psmallmatrix}$   &  $[v_{10},v_{11}]$ \\
$3i+2$ &  $0\rightarrow 0$  & $\mu$  & $[x_1,x_1]$ & $\begin{psmallmatrix}
    0 & 0  \\
    1& 1
\end{psmallmatrix}$   &  $[v_{01},v_{01}]$ \\
\end{tabular}
 \captionof{table}{Stochastic Hybrid System description}
 \label{trans}
\end{center}
We provide in the following a detailed explanation to each transition reported in Table \ref{trans}:
\begin{enumerate}
\item The set of transitions from $l=1$ till $l=i+1$ represents a new packet arrival of the stream of interest. As explained in our system model section, a new packet arrival will replace the packet in the waiting room. In the case where a packet of the stream of interest is already being served by the server, the newly arrived packet will preempt its service and take its place. We can therefore see that this transition will have no effect on the age of this stream at the monitor $x_0$. However, the age of the system's packet $x_1$ falls to $0$.
\item The transitions set spanning from $l=i+2$ till $l=2i+1$ corresponds to packets arrivals from higher priority streams. In state $q=0$, a packet of the stream of interest (either real or fake) is being served. As there are $i$ higher priority streams, any arrival from \emph{either} of them will preempt the service of this packet and the packet of the stream of interest is brought back to its own waiting room. This transition has a rate of $i\lambda$. Now, in state $q=1$, there is already one higher priority packet being served (let's suppose it belongs to stream $k'$); we do not care about its exact priority as we only care that it is of higher priority than the stream of interest. A new arrival for stream $k'$ will replace the packet currently in service but will not have any effect on the system from the point of view of the stream of interest as there was already a packet for stream $k'$ being served. Therefore, this transition is omitted. However, a new packet arrival of the $i-1$ remaining higher priority streams will have an effect on the system as there would be now $2$ packets to be served ahead of the packet of the stream of interest. This transition has a rate of $(i-1)\lambda$ and will take the system from $q_l=1$ to $q'_l=2$. The same reasoning goes on till the last transition $l=2i+1$. All these transitions will have no effect on the age process of the stream of interest and therefore $\boldsymbol{A_l}=\boldsymbol{I}$.
\item The transitions spanning from $2i+2$ till $3i+1$ corresponds to the server finishing the transmission of a packet of higher priority streams. These transitions have a rate of $\mu$ and will have no effect on the age process of the stream of interest (i.e. $\boldsymbol{A_l}=\boldsymbol{I}$). The last transition $l=3i+2$ takes place when a packet of the stream of interest finishes being served. This will reset the age at the monitor to that of the delivered packet $x'_0=x_1$. A fake update is then generated with the same age as the previously transmitted one $x'_1=x_1$.
\end{enumerate}
As for the differential equations that portray the evolution of the age process in each discrete state, we have that in each state $q\in\mathbb{Q}$, $x_{0}(t)$ and $x_{1}(t)$ increase at a unit rate:
\begin{equation}
\boldsymbol{b}_q=[1\:\:1] \quad\forall q\in \mathbb{Q}
\label{bstuff}
\end{equation}
\begin{figure}[!ht]
\centering
\includegraphics[width=.9\linewidth]{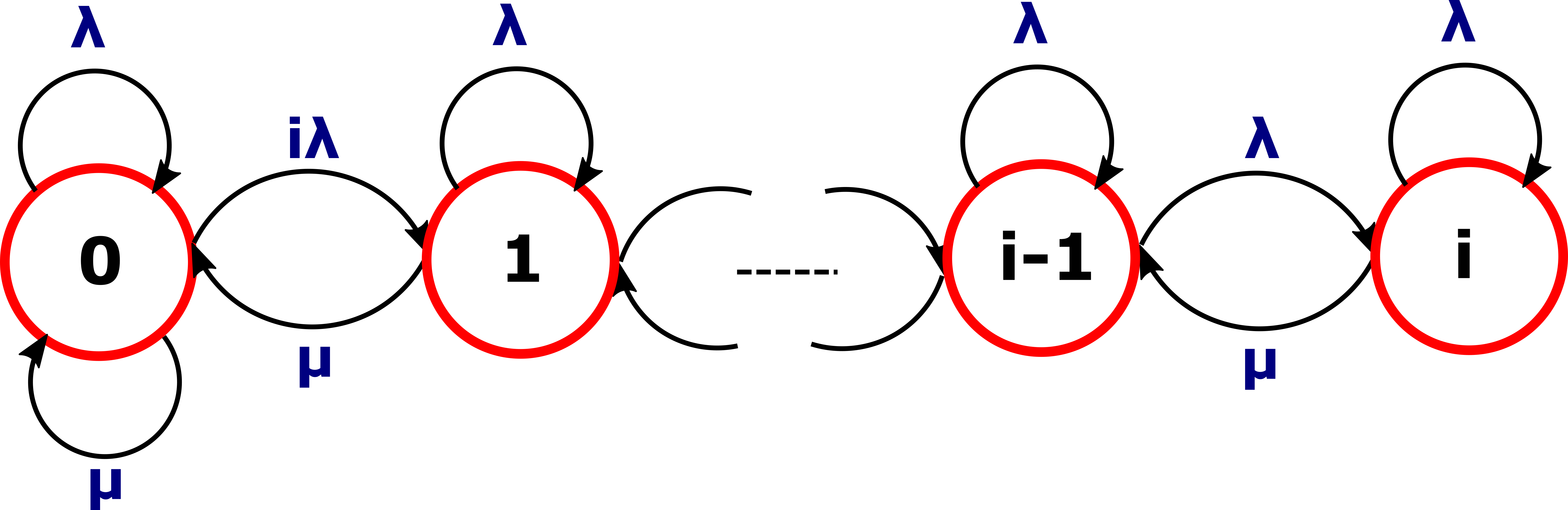}
\setlength{\belowcaptionskip}{-10pt}
\caption{Illustration of the stochastic hybrid systems Markov chain}
\label{effect}
\end{figure}

To be able to apply Theorem \ref{theomhem}, we start by investigating the stationary distribution of the Markov Chain that models the
transitions reported in Table \ref{trans}. To do so, we provide the following proposition.
\begin{proposition} The continuous time Markov chain is irreducible, time-reversible and admits $\overline{\pi}(k;\lambda,\mu)$ as stationary distribution for any  state $0\leq k\leq i$ where:
\begin{equation}
\overline{\pi}_k=\frac{\lambda^ki!}{\mu^k(i-k)!}\overline{\pi}_0
\label{Stationary1}
\end{equation}
and:
\begin{equation}
\overline{\pi}_0=\frac{1}{\sum\limits_{k=0}^{i}\frac{\lambda^ki!}{\mu^k(i-k)!}}
\end{equation}
\label{stationarydist}
\end{proposition}
\begin{IEEEproof}
We proceed to prove the proposition by induction:
\begin{itemize}
\item[--] For $k=1$, we have from eq. (\ref{Stationary1}) that $\overline{\pi}_1=\frac{\lambda i}{\mu}\overline{\pi}_0$. By formulating the general balance equation at state $k=0$, we have that $\overline{\pi}_0(\lambda i)=\mu\overline{\pi}_1$ and the proposition is therefore true for $k=1$.
\item[--] We suppose that the proposition is true up to $k\leq i-1$ and we formulate the general balance equation at state $k$:
\begin{equation}
\overline{\pi}_{k}(\mu+\lambda(i-k))=\lambda(i-k+1)\overline{\pi}_{k-1}+\mu\overline{\pi}_{k+1}
\end{equation}
By substituting $\overline{\pi}_{k}$ and $\overline{\pi}_{k-1}$ by their supposed values, we can verify that $\overline{\pi}_{k+1}=\frac{\lambda^{k+1}i!}{\mu^{k+1}(i-(k+1))!}\overline{\pi}_0$ which concludes our proof. The time-reversibility can be easily verified by showing that the stationary distribution provided in eq. (\ref{Stationary1}) satisfies the detailed balance equations.
\end{itemize}
\end{IEEEproof}
Armed with this proposition, we can proceed to calculate the average age for the stream of interest $k$. On this note, we provide the following theorem.
\begin{theorem}
The average age of a stream of interest having $i$ streams with higher priority above it is\footnote{One can notice that when $i=0$, we have $\overline{\Delta}_0=\frac{1}{\lambda}+\frac{1}{\mu}$; the expression coincides with that of an M/M/1/1 system with preemption reported in \cite{DBLP:journals/corr/YatesK16}. This is in accordance with the fact that the stream with the highest priority is not affected by any other stream}:
\begin{equation}
\overline{\Delta}^{WQ}_i=\overline{v}_{00}+\sum\limits_{k=1}^{i}\overline{v}_{k0}
\end{equation}
where:
\begin{equation}
\overline{v}_{00}=\frac{1}{\mu}+\sum\limits_{j=0}^{i}\frac{\lambda^ji!}{(i-j)!\sum\limits_{k=0}^{i}\frac{\lambda^ki!}{\mu^k(i-k)!}\displaystyle\prod_{h=0}^{j} a_{h}}
\label{v00finalbefore}
\end{equation}
with $a_h\in\mathbb{R}$ is a real sequence defined as:
\begin{equation*}
a_h=
\begin{cases}
\lambda+\mu  \qquad\qquad\qquad\qquad\qquad\qquad\qquad h=i\neq0\\
(i-h+1)\lambda+\mu-\frac{(i-h)\lambda\mu}{a_{h+1}} \qquad\qquad 1\leq h\leq i-1\\
(i+1)\lambda-\frac{i\lambda\mu}{a_{1}} \qquad\qquad\qquad\qquad\qquad\qquad h=0
\end{cases}
\label{Stationary}
\end{equation*}
and:
\begin{equation}
\overline{v}_{k0}=\frac{(i-k+1)\lambda}{\mu}\overline{v}_{(k-1)0}+\sum\limits_{j=k}^{i}\frac{\lambda^ji!}{\mu(i-j)!\sum\limits_{k=0}^{i}\frac{\lambda^ki!}{\mu^k(i-k)!}}\quad 1\leq k \leq i
\label{general0before}
\end{equation}
\label{theoremaverage}
\end{theorem}
\begin{IEEEproof}
With the stationary distribution of the Markov chain being found, we start by applying Theorem \ref{theomhem}. To properly find the vectors $\overline{\boldsymbol{v}}_{q}=[\overline{v}_{q0},\overline{v}_{q1}]\:\: \forall q\in\mathbb{Q}$, we proceed by applying the aforementioned theorem in the states where the service transitions with rate $\mu$ have no effect on the age process $\boldsymbol{x}(t)$ (i.e. $\boldsymbol{A_l}=\boldsymbol{I}$). More specifically, we start by applying Theorem \ref{theomhem} in the state $q=i$ and go backwards in the discrete state space. By doing so and by focusing on the first component of the vector $\overline{\boldsymbol{v}}_{i}$, we end up with the following:
\begin{equation}
\overline{v}_{i0}=\frac{\overline{\pi}_i}{\mu}+\frac{\lambda}{\mu}\overline{v}_{(i-1)0}
\end{equation} 
By doing a successive backwards induction till state $k=1$, we can verify that for all states $1\leq k\leq i$, we have:
\begin{equation}
\overline{v}_{k0}=\sum\limits_{j=k}^{i}\frac{\overline{\pi}_{j}}{\mu}+\frac{(i-k+1)\lambda}{\mu}\overline{v}_{(k-1)0}
\label{general0}
\end{equation}
The next step consists of formulating the results of Theorem \ref{theomhem} for the second component of the vector $\overline{\boldsymbol{v}}_{q}$. The tricky part with this formulation is the fact that the denominator change in each state. To highlight this, we can see that:
\begin{equation}
\overline{v}_{i1}=\frac{\overline{\pi}_i}{\mu+\lambda}+\frac{\lambda}{\mu+\lambda}\overline{v}_{(i-1)1}
\label{demo1}
\end{equation} 
In the state $k=i-1$, and by taking into account the previous equation (\ref{demo1}), we end up with:
\begin{multline}
\overline{v}_{(i-1)1}=\frac{\mu\overline{\pi}_i}{(\mu+\lambda)(2\lambda+\mu-\frac{\lambda\mu}{\lambda+\mu})}+\frac{\overline{\pi}_{i-1}}{2\lambda+\mu-\frac{\lambda\mu}{\lambda+\mu}}\\+\frac{2\lambda}{2\lambda+\mu-\frac{\lambda\mu}{\lambda+\mu}}\overline{v}_{(i-2)1}
\end{multline}
Therefore, we carefully employ a backwards induction to conclude the following closed form for all states $1\leq k\leq i$:
\begin{equation}
\overline{v}_{k1}=\sum\limits_{j=k}^{i}\frac{\mu^{j-k}\overline{\pi}_j}{\displaystyle\prod_{h=0}^{j-k} a_{k+h}}+\frac{(i-k+1)\lambda}{a_k}\overline{v}_{(k-1)1}
\label{general1}
\end{equation}
where $a_k\in\mathbb{R}$ is a real sequence that is defined for $1\leq k\leq i$ as follows:
\begin{equation*}
a_{i}=\lambda+\mu
\end{equation*}
\begin{equation}
a_{k}=(i-k+1)\lambda+\mu-\frac{(i-k)\lambda\mu}{a_{k+1}} \quad 1\leq k\leq i-1
\end{equation}
With the general expression of both $\overline{v}_{k0}$ and $\overline{v}_{k1}$ for all $1\leq k\leq i$ at our disposal, we continue by formulating the equations at state $k=0$ where the transition of rate $\mu$ will induce resets in the age processes:
\begin{equation}
\overline{v}_{00}(i\lambda+\mu)=\overline{\pi}_0+\mu \overline{v}_{01}+\mu \overline{v}_{10}
\end{equation}
By using the general expression in eq. (\ref{general0}), we can conclude that:
\begin{equation}
\overline{v}_{00}=\frac{1}{\mu}+\overline{v}_{01}
\end{equation}
As the goal is to find $\sum\limits_{k=0}^{i}\overline{v}_{k0}$, we proceed with calculating $\overline{v}_{01}$:
\begin{equation}
\overline{v}_{01}((i+1)\lambda)=\overline{\pi}_0+\mu \overline{v}_{11}
\end{equation}
By using the general expression in eq. (\ref{general1}), we can conclude that:
\begin{equation}
\overline{v}_{01}=\sum\limits_{j=0}^{i}\frac{\mu^{j}\overline{\pi}_j}{\displaystyle\prod_{h=0}^{j} a_{h}}
\end{equation}
where $a_0=(i+1)\lambda-\frac{i\lambda\mu}{a_{1}}$. All in all, we can conclude that:
\begin{equation}
\overline{v}_{00}=\frac{1}{\mu}+\sum\limits_{j=0}^{i}\frac{\mu^{j}\overline{\pi}_j}{\displaystyle\prod_{h=0}^{j} a_{h}}=\frac{1}{\mu}+\sum\limits_{j=0}^{i}\frac{\lambda^ji!}{(i-j)!\sum\limits_{k=0}^{i}\frac{\lambda^ki!}{\mu^k(i-k)!}\displaystyle\prod_{h=0}^{j} a_{h}}
\label{v00final}
\end{equation}
Knowing that $\overline{\Delta}^{WQ}_i=\sum\limits_{k=0}^{i}\overline{v}_{k0}$ and by taking into account the results of eq. (\ref{general0}) and (\ref{v00final}), we conclude our proof.
%
%
It is worth mentioning that, as the results of this theorem are general for any priority level $i$, the total average age of the system can be easily calculated as follows: $\overline{\Delta}^{WQ}=\sum\limits_{i=0}^{N-1}\overline{\Delta}^{WQ}_i$. 
\end{IEEEproof}
\section{Numerical Results}
\begin{figure*}[ht]
\centering
\begin{subfigure}{0.25\textwidth}
  \centering
  \includegraphics[width=.9\linewidth]{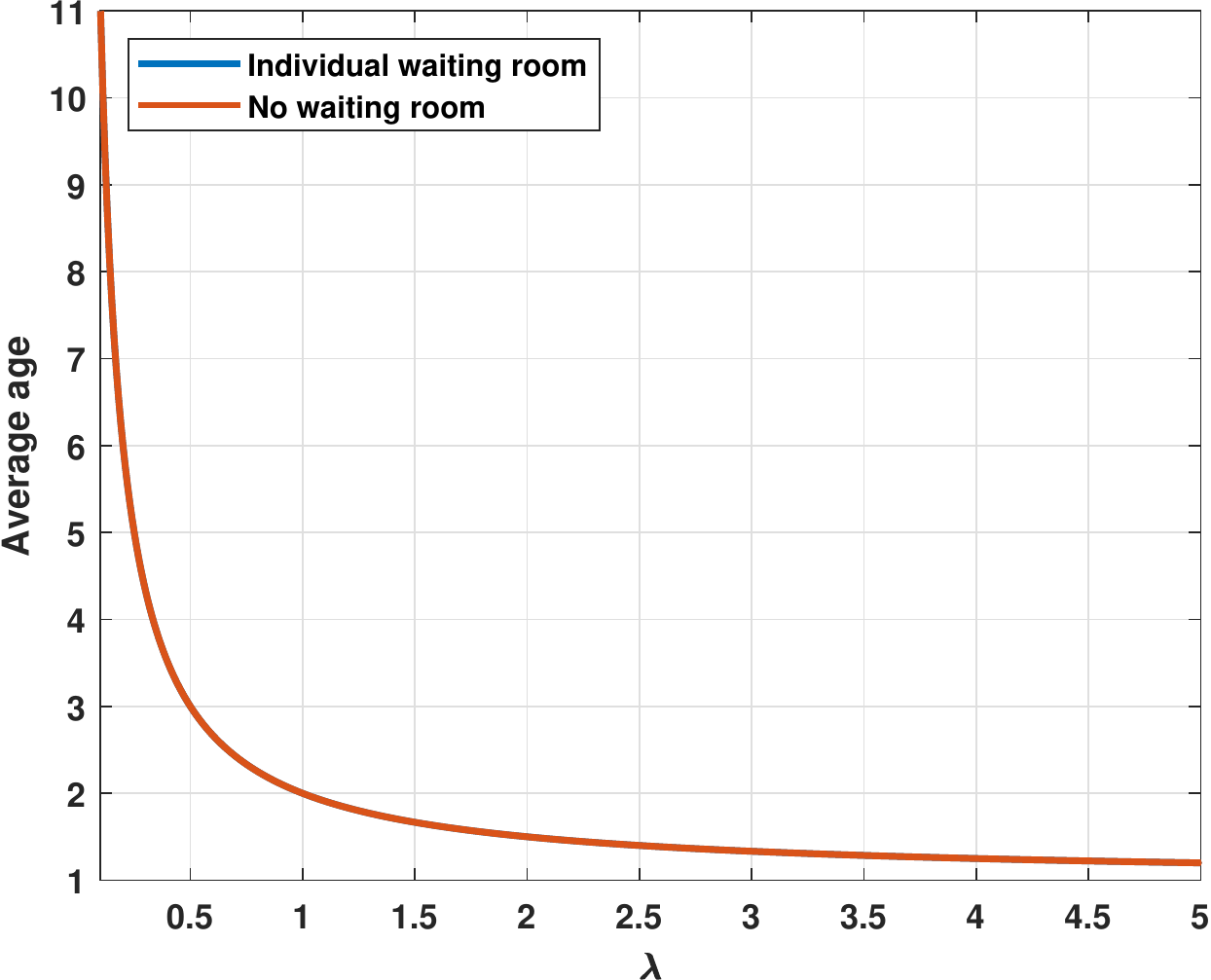}
  \caption{Stream $1$}
    \label{sim1}
\end{subfigure}%
\begin{subfigure}{0.25\textwidth}
\centering
  \includegraphics[width=.9\linewidth]{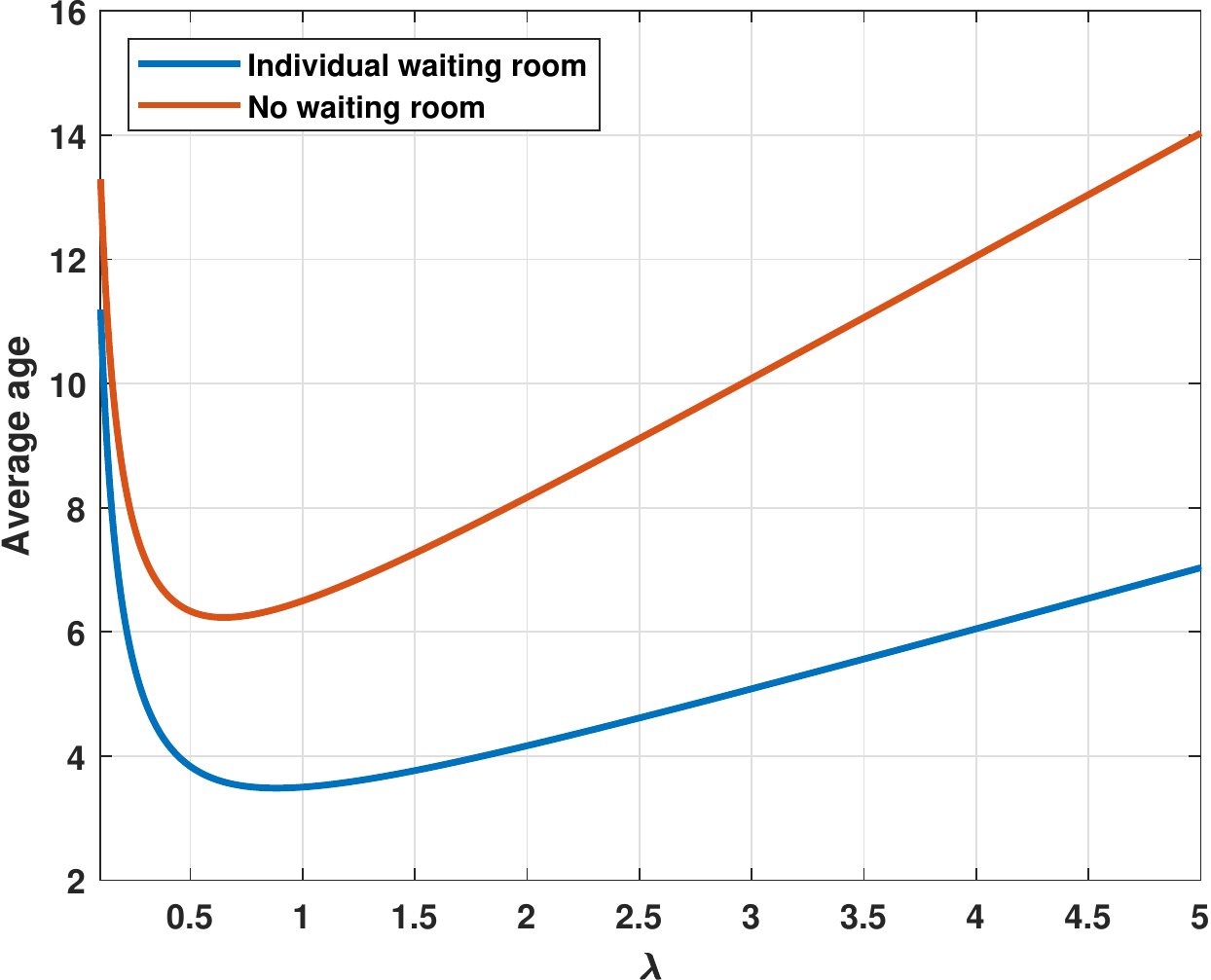}
  \caption{Stream $2$}
\label{sim2}
\end{subfigure}%
\begin{subfigure}{0.25\textwidth}
\centering
  \includegraphics[width=.9\linewidth]{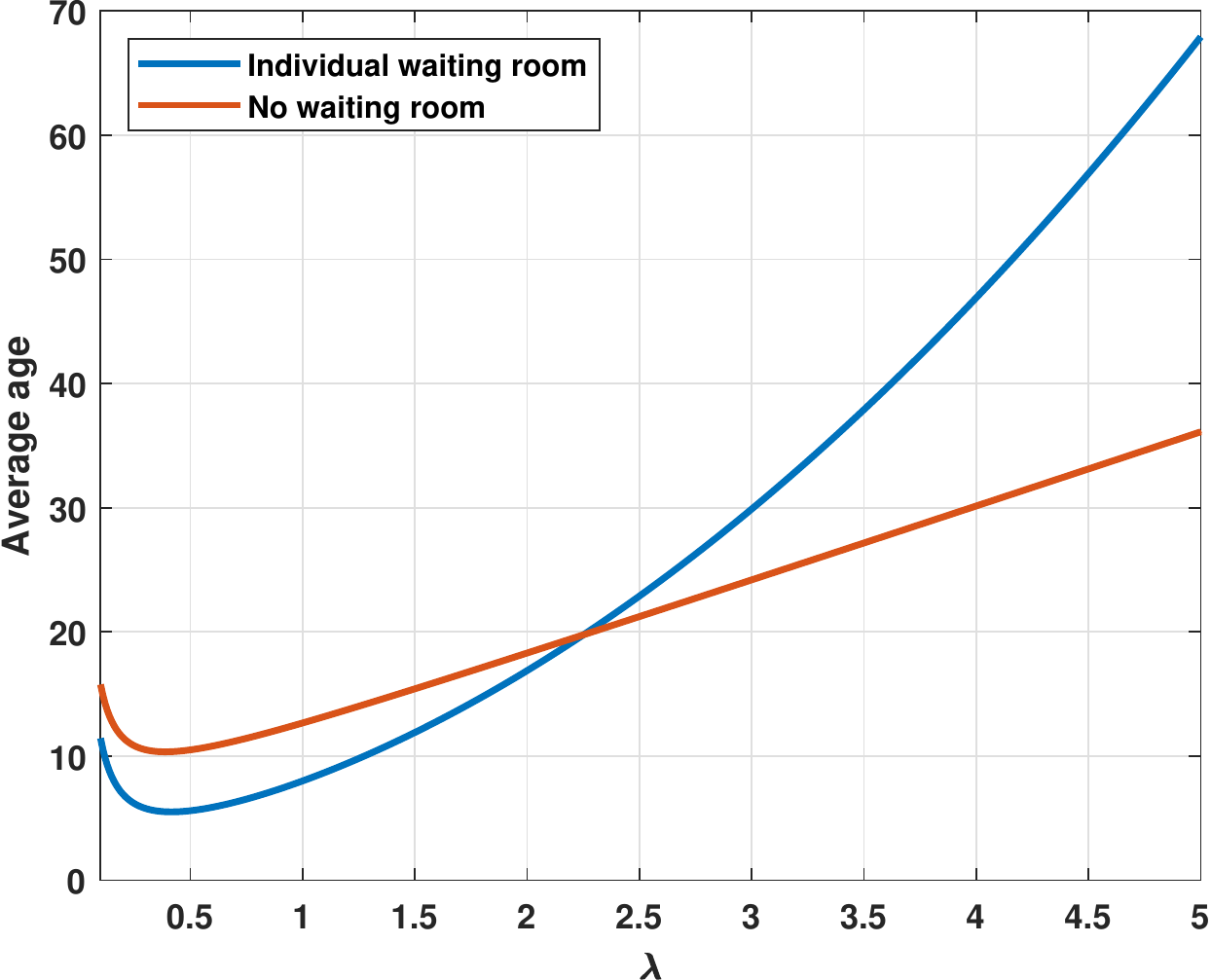}
  \caption{Stream $3$}
\label{sim3}
\end{subfigure}%
\begin{subfigure}{0.25\textwidth}
  \centering
  \includegraphics[width=.9\linewidth]{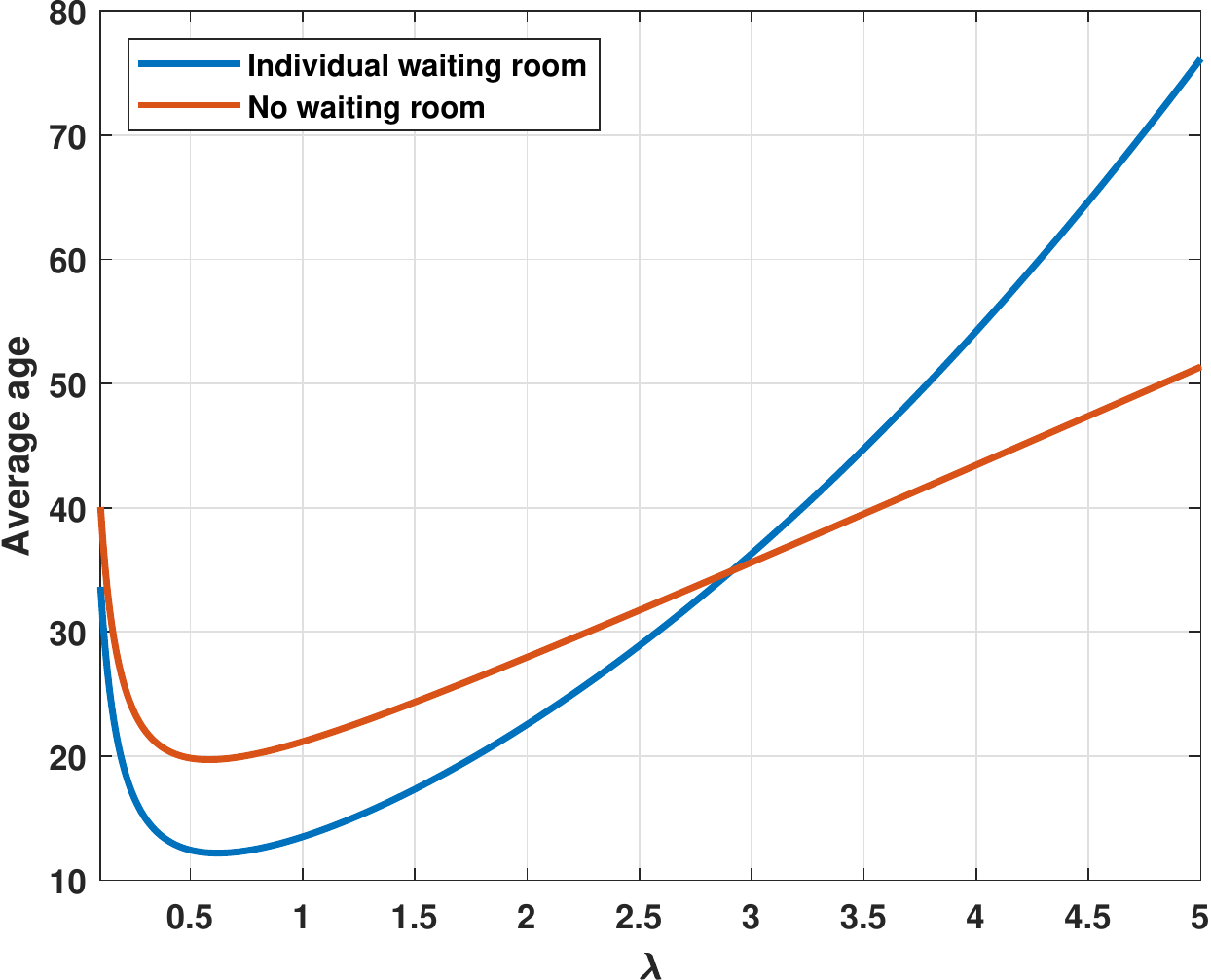}
  \caption{Total streams}
\label{sim4}
\end{subfigure}%
\caption{The average age of each stream in function of the arrival rate $\lambda$}
\vspace{-20pt}
\end{figure*}
The main goal is to compare the average age of each stream, along with the total average age, in the case where a waiting room exists for each stream (referred to as $WQ$) and the latter case where there is no waiting room (referred to as $NQ$). The average age in the latter can be found in \cite{8437591}. We consider in the following that $\mu=1$ and $N=3$. We can see in Fig. \ref{sim1} that stream $1$ in both cases achieves the same average age since, in both cases, the highest priority stream sees the system as a preemptive M/M/1/1 system. For the second stream, as seen in Fig. \ref{sim2}, having a waiting room clearly helps $\forall \lambda$. This is due to the fact that if a packet of stream $2$ is being served when a packet of stream $1$ arrives, it is not discarded after preemption and is therefore resumed right after. An interesting observation can be seen in Fig. \ref{sim3} for stream $3$: although a preempted packet of stream $3$ is not discarded, yet this stream only exhibits an advantage with respect to $NQ$ for low $\lambda$. As $\lambda$ increases, the gap get smaller until both curves intersect and having a waiting room worsen its performance in terms of average age. The reason behind this is the fact that, although the packets of stream $3$ are not being discarded after preemption, the same thing is happening to stream $2$. Therefore, packets of stream $3$ have to wait for the preempted packet of stream $2$, by stream $1$, to continue service before it can actually be served itself. Due to this observation, we can see in Fig. \ref{sim4} that the total average age with waiting room is smaller than its counterpart when $\lambda<\lambda_{PASS}$ and is higher otherwise. $\lambda_{PASS}$ denotes the arrival rate corresponding to the intersection of the two curves. The exact value of $\lambda_{PASS}$ can be theoretically found simply by using the results of Theorem \ref{theoremaverage} in our paper and that of \cite{8437591}.

To further highlight this trade-off, we vary the number of streams $N$ and report the results in the following table. The first thing we can notice is that in both cases, the optimum is attained for the same arrival rate $\lambda_{OPT}$. However, $\overline{\Delta}^{WQ}_{OPT}$ is always smaller than $\overline{\Delta}^{NQ}_{OPT}$ with the gap between them increasing as $N$ grows. It is worth noting that the optimum is achieved for smaller values of $\lambda$ as $N$ grows higher due to the congestion. As for the trade-off, we can see that, as $N$ grows, $\lambda_{PASS}$ becomes smaller and buffering packets instead of discarding them after preemption worsen the performance in terms of age when $\lambda>\lambda_{PASS}$. Based on the results reported in the present work, we can have the following conclusions:
\begin{enumerate}
\item if the device has control over the packets generation rate $\lambda$, it is always better to opt for having a buffer space for each stream since it was shown that this always outperforms the no buffer case for $\lambda=\lambda_{OPT}$.
\item in the latter case where the device does not have such control, the decision depends on the arrival rate itself: if it is low (more specifically, below $\lambda_{PASS}$), it is better to keep a buffer for each stream while forgoing buffering would achieve better age performance when $\lambda$ surpasses this critical value.
\end{enumerate}
\def\arraystretch{1.5}
\begin{center}
\begin{tabular}{|c|c|c|c|c|c|}
 \hline
 $N$ & $\overline{\Delta}^{WQ}_{OPT}$  &  $\lambda^{WQ}_{OPT}$ & $\overline{\Delta}^{NQ}_{OPT}$ & $\lambda^{NQ}_{OPT}$ & $\lambda_{PASS}$ \\
  \hline
 $3$  & $12.18$ & $0.62$  & $19.71$& $0.62$ & $2.92$ \\
 $5$  & $33$ & $0.3$  & $55$& $0.3$ & $0.7$ \\
 $8$  & $81.7$ & $0.16$  & $140$& $0.16$ & $0.31$ \\
 \hline
\end{tabular}
 \captionof{table}{Comparison between $WQ$ and $NQ$ scenarios}
 \label{valuesktir}
\end{center}

\section{Conclusion}
In this paper, we have investigated the scenario where $N$ information streams, each with different priority and own waiting room, share a common server. Using SHS tools, we were able to find a closed form of the average AoI of each stream. Armed with this, we were able to highlight an interesting observation: in some cases, having no waiting room can be beneficial to the age performance. These specific cases were thoroughly discussed and numerical results that corroborate these findings were provided.
\bibliographystyle{IEEEtran}
\bibliography{trialout}

\end{document}